\documentclass[fleqn,12pt]{wlscirep}
\usepackage{amsmath}
\usepackage{graphicx}
\usepackage{footmisc}
\usepackage{color}

\begin{document}
\title {Composite Hofstadter bands with Dirac fermion spectrum of fractional quantum Hall states}
\author[1]{Igor N. Karnaukhov}
\affil[1]{G.V. Kurdyumov Institute for Metal Physics, 36 Vernadsky Boulevard, 03142 Kyiv, Ukraine}
\affil[*]{karnaui@yahoo.com}
\begin{abstract}
The  fractional quantum Hall effect (FQHE) is studied in the semiclassical limit in the framework of the Hofstadter model with a short-range interaction between fermions. In the mean-field approximation, the repulsion between fermions leads to a periodic potential. Numerical calculations show that in the case of the periodic potential with a period that is a multiple on $\frac{1}{\nu}$ of the magnetic cell ($\nu$ is filling of a separated band) composite Hofstadter bands (HBs) are formed.  The composite HBs are split into $\frac{1}{\nu}$ subbands, which are separated by the Dirac points. The Chern number of $\gamma$-full filled composite HBs is equal to the Chern number  $C_{\gamma}$ of the corresponding HB.  The Chern number, equal to $\nu C_{\gamma}$, corresponds to $\nu$-filling of $\gamma$-composite HB.  Thus, FQHE is realized by fractional filling of composite HBs. 
\end{abstract}
\maketitle
\section*{Introduction}

The explanation of a wide range of experimental results based on the Hofstadter model \cite{Har,Hof} model makes it significant not only in theoretical but also in mathematical physics. This is due, first of all, to its successful application in the study of topological states of two-dimensional electron liquid. First of all, we should note the study within the framework of this model of the integer and fractional quantum Hall effects experimentally observed in two-dimensional electron liquid in strong magnetic fields \cite{2}.

A composite fermion paradigm \cite{Jain} describes FQHE \cite{2} quite well and explains the experimentally observed series of quantum Hall plateaus. 
For half-filled HB, that corresponds to $\nu=\frac{1}{2}$ FQHE \cite{CF1}, in the zero bare mass limit of composite fermions particle-hole symmetry is exact. In this case an effective field theory for such a fermion has been proposed \cite{Son}.
It is clear that the case $\nu=\frac{1}{2}$  is unique in the sense of symmetry.  Nevertheless, the conclusion about the Dirac behavior of the spectrum of the electron liquid in the state corresponding to FQHE  is not only new but also very interesting \cite{Son,CF2,CF3}. 
It is not clear how exactly the half-filling should stand out against the background of other quantum Hall plateaus.
The experiment dictates rather rigid conditions, namely: to describe within one theory a set of sequences-series of quantum Hall plateaus at arbitrary filling (integer and fractional). 

Despite the attractiveness of the Hofstadter model, it is unable to explain FQHE because in this case it is necessary to explicitly  account for the interaction between electrons (the single-particle Hofstadter model explains only the integer quantum Hall effect). In the simplest case we can speak about a periodic potential, which takes into account the interactions between electrons in the mean-field approximation \cite{0,IK0}.  Despite the fact that in \cite{0,IK0} almost identical models are considered, the calculation results are opposite, namely, according to \cite{0} the periodic potential cannot explain the state with fractional conductance, whereas in \cite{IK0} it is shown that the states of electron liquid in the periodic potential can lead to FQHE. In \cite{IK0} in particular it was shown that in the case of less than half-filling of HBs a stable structure with a period  $a=\frac{1}{\nu} q$ (multiple of the magnetic sublattice cell $q$) is realized. The results of calculations are obtained for $q>>1$. Experimentally realized magnetic fields correspond to magnetic sublattices with $q>>1$, so it makes sense to consider solution of the problem in semiclassical limit. Note, that topological invariants of the electron liquid  not depend on the magnitudes of magnetic field and hopping integrals, i.e. it is universal in this sense. 

The aim of the paper is to study FQHE in the framework of the Hofstadter model, in which the interaction of fermions is reduced to a periodic potential. In this case it is better to operate not with composite fermions but with composite HBs, which are realized as a result of this interaction. 
Composite HB is split into $\frac{1}{\nu}$-subbands by the Dirac points forming its fine structure. The topological order of the subbands is determined by fractional Chern numbers, which correspond to the FQHE states.


\section*{The model}

We will analyze a two-dimensional fermion liquid in a transverse magnetic field $H$,  the Hamiltonian of which is determined in the framework of the Hofstadter Hamiltonian ${\cal H} ={\cal H}_0+{\cal H}_{int}$ \cite{Har,Hof}
\begin{eqnarray}
&& {\cal H}_0 =\sum_{m,n}[ t_x (m,n) a^\dagger_{m,n}a_{m+1,n}+t_y (m,n) a^\dagger_{m,n}a_{m,n+1} + H.c.]
- \mu \sum_{{\textbf{j}}} n_{{\textbf{j}}},
\label{eq-1}
\end{eqnarray}
\begin{eqnarray}
 && {\cal H}_{int}=U \sum_{{\textbf{j}}} n_{{\textbf{j}}}n_{{\textbf{j+1}}},
    \label{eq-2}
\end{eqnarray}
where $a^\dagger_{\textbf{j}} $ and $a_{\textbf{j}}$ are the operators of spinless fermions at a site ${\textbf{j}}=\{m,n\}$, $n_{{\textbf{j}}}=a^\dagger_{{\textbf{j}}}a _{{\textbf{j}}}$ is the density operators, $\mu$ is the chemical potential,
The  Hamiltonian  (1) describes the nearest-neighbor hoppings of fermions with
different hopping integrals along the $x$-direction $t_x({\textbf{j}}) = 1$ and the $y$-direction $t_ y({\textbf{j}}) = \exp[2 i \pi m \phi]$. A magnetic flux through a unit cell $\phi = \frac{H }{ \Phi_0}$ is determined in the quantum flux unit ${\Phi_0=h/e}$, a homogeneous magnetic field $H$ is defined by the vector potential $A_y=H x$ which is directed along $y$-direction. ${\cal H}_{int}$ term is determined by an interaction strength $U$.

We consider the two-dimensional electron system in the stripe geometry with open boundary conditions for the boundaries along the $y$-direction with  linear size L. The magnetic field enters the Hamiltonian (\ref{eq-1}) in the form of the magnetic flux $\phi$ through the unit cell.

\section*{Solution of the problem}

On the one hand, the Chern number characterizes the topological order of an isolated band, on the other hand, it determines the Hall conductance, as a result of which it relates the Hall conductance to the topology of the electronic liquid. The Chern number clearly does not depend on the bare values of the hopping integrals and magnitude of magnetic field. It is sufficient to calculate the Hall conductance at certain values of the parameters of the model Hamiltonian (1),(2), which do not change the topology of the object, without investigating the whole region of their ranging.  
In the case of experimentally realizable magnetic fields, that corresponds to $q \sim 10^3 -10^4$, it makes sense to consider the quantum Hall effect in the semiclassical limit $q>>1$ for an arbitrary rational magnetic flux $\phi=\frac{p}{q}$ (here $p$ and $q$ are prime integers). Without loss of generality of the problem, we consider the case $p=1$.

Taking into account $n_{\textbf{k}}=\frac{1}{V}\sum_j exp(i\textbf{k j})n_{\textbf{j}}$ an interaction term (2) can be conveniently redefined in the momentum representation
${\cal H}_{int} = V U  \sum_{\textbf{k}}\cos{\textbf{k}}n_{\textbf{k}} n_{-\textbf{k}}$, the volume is equal to $V=L \times L$. In the mean field approach, we rewrite this term as follows ${\cal H}_{int} =V\sum_{\textbf{k}}\cos{\textbf{k}}(\lambda_{\textbf{k}}n_{\textbf{-k}}+\lambda_{\textbf{-k}}n_{\textbf{k}})$ with an effective field $\lambda_{\textbf{K}}=U <n_{\textbf{k}}>$, which is determined by a free parameter-wave vector $\textbf{k}$, the value of which corresponds to energy minimum.
In the semiclassical limit a magnetic scale is large $q>>1$, which corresponds to small values $k\sim 10^{-3}-10^{-4}$, the density of fermions $\rho$ for the states near the low energy edge of  the spectrum is small $\sim \frac{1}{q}$.
In the small k-limit the expression for $\lambda_{\textbf{k}}$ is simplified  $\lambda_{\textbf{k}}=\lambda +0(k^2)$, where $\lambda =2 U\rho$ and 
\begin{eqnarray}
&&{\cal H}_{eff} =\lambda\sum_{\textbf{j}}\cos(\textbf{k j})n_\textbf{j}.
\label{eq-3}
\end{eqnarray}
 
In the semiclassical limit the  spectrum of quasi-particle excitations is defined by quasi-flat bands  (HBs) separated by gaps (the widths of these bands are much smaller than the gaps between them). The wave function has different behavior in the $y$- and $x$-directions. In the $y$-direction, the lattice decouples into chains in which the fermions are free ($m$ is a coordinate of the $y-$chain in the x-direction), while in the $x$-direction the wave function has a localized behavior with overlap between chains. The  chains are connected by single-particle tunneling with tunneling constant $t_x$. The states with different $m$ are bounded via a magnetic flux and form  a magnetic cell.
The effective hopping integrals in the $x-$ and $y-$ directions $\delta t_x,\delta t_y$, which determine the width of the quasi-flat HB, differ in magnitude by several orders of magnitude ($\delta t_x<<\delta t_y$), so the band width is determined by the integral of fermionic hopping along the $y$-direction. In this case the repulsion along the $x$-direction dominates, since the effective interaction between fermions is defined with respect to the hopping integrals.

We study the 2D system in a hollow cylindrical geometry with open boundary conditions (a cylinder axis along the $x$-direction and the boundaries along the $y$-direction). The interaction (\ref{eq-2}) does not break the time reversal symmetry of the model Hamiltonian (\ref{eq-1})-(\ref{eq-2}), the effective Hamiltonian (\ref{eq-3}) should also not break these symmetries for rational fluxes. These conditions are fulfilled in the case when $\textbf{k}=(k,0)$, where $k$ and $q$ form the states with rational periods. Making the Ansatz for the wave function $\psi(m,n)=\exp(i k_y n)g_m$ (which determines the state with the energy $\epsilon$) we obtain the Harper equation for the Hamiltonian (\ref{eq-1}), (\ref{eq-3})
\begin{equation}
\epsilon g_m =-g_{m+1}-g_{m-1}-2\cos(k_y +2\pi m \phi)g_m +\lambda \cos(k m)g_m.
 \label{eq-Har}
\end{equation}
This equation is studied in \cite{IK0}. 

\begin{figure}[tp]
     \centering{\leavevmode}
     \begin{minipage}[h]{.4915\linewidth}
\center{\includegraphics[width=\linewidth]{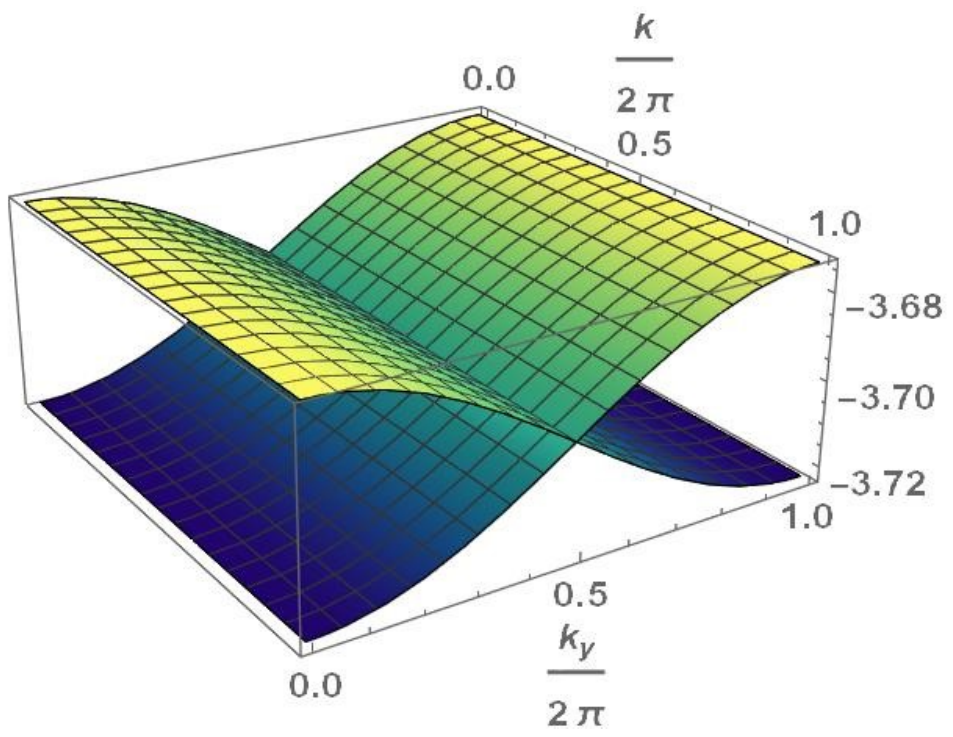} a)\\
               }
    \end{minipage}
     \centering{\leavevmode}
\begin{minipage}[h]{.4915\linewidth}
\center{\includegraphics[width=\linewidth]{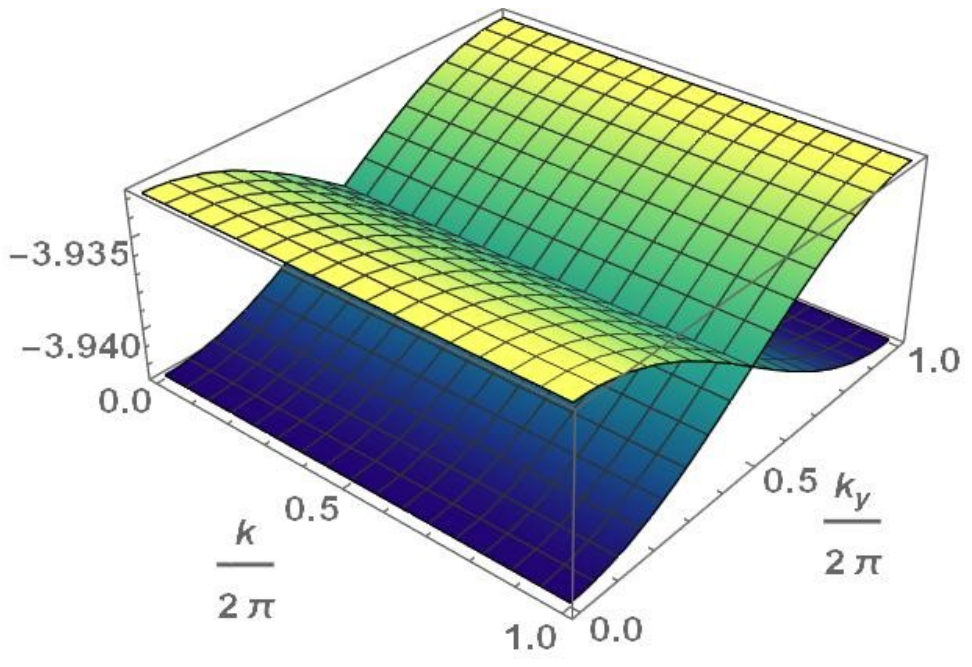} b)\\
                }
    \end{minipage}
\centering{\leavevmode}
\begin{minipage}[h]{.35\linewidth}
\center{\includegraphics[width=\linewidth]{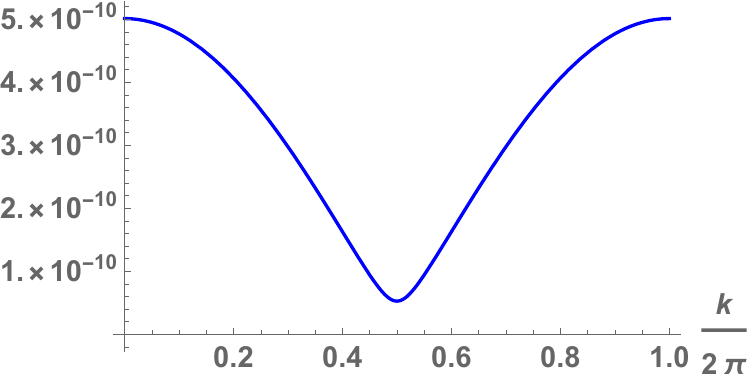} c)\\
             }
    \end{minipage}
    \centering{\leavevmode}
\begin{minipage}[h]{.35\linewidth}
\center{\includegraphics[width=\linewidth]{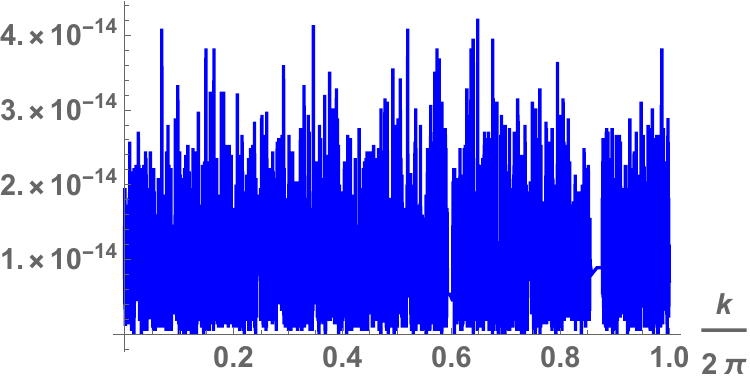} d)\\
             }
    \end{minipage}
\caption{(Color online)\\$\frac{1}{2}$-filling ($a=2q$):
the spectrum of the first (lowest energy) HB as a function of wave vector $(k,k_y)$, calculated at $U=\frac{1}{2}$, $\rho=\frac{1}{a}$: a) $q=20$, b) $q=100$; the quasi-gap in the spectrum of the first HB as a function wave vector $k$ calculated at  $k_y=\pi$ c) $q=20$, d) $q=100$.
}
\label{fig:1}
\end{figure}
\begin{figure}[tp]
     \centering{\leavevmode}
     \begin{minipage}[h]{.4915\linewidth}
\center{\includegraphics[width=\linewidth]{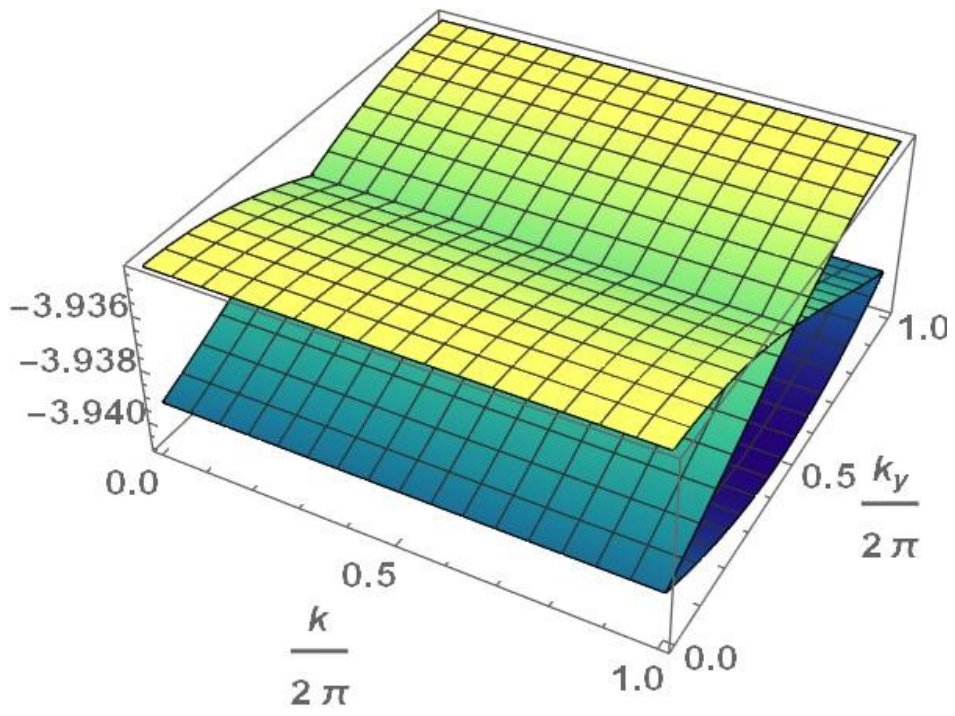} a)\\
               }
    \end{minipage}
     \centering{\leavevmode}
\begin{minipage}[h]{.4915\linewidth}
\center{\includegraphics[width=\linewidth]{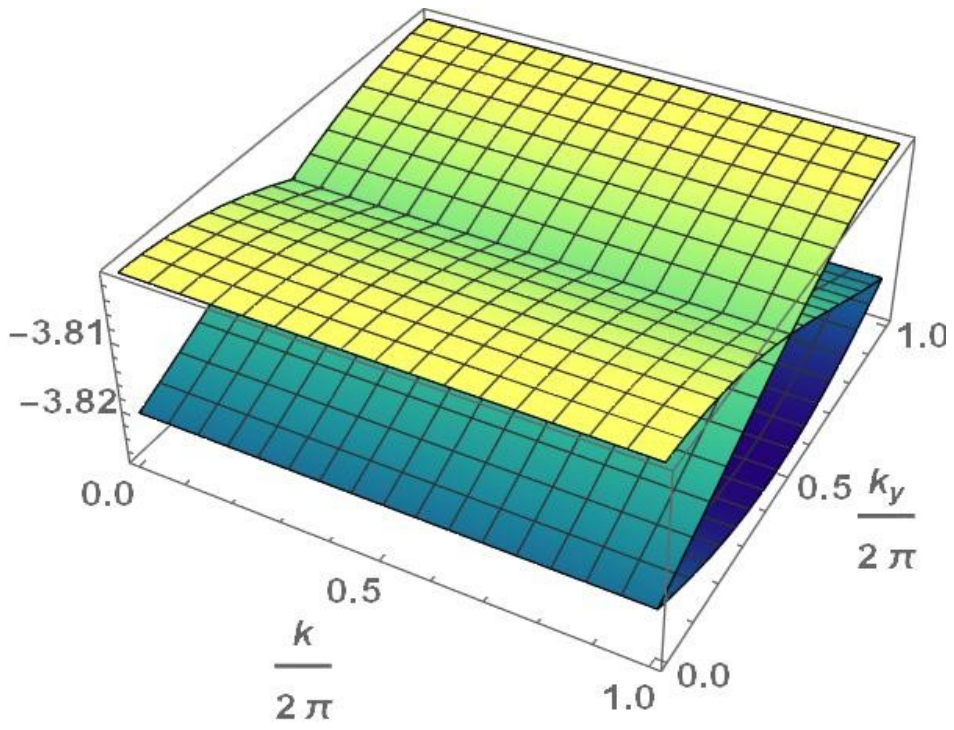} b)\\
                }
    \end{minipage}
\caption{(Color online)\\
$\frac{1}{3}$-filling ($a=3 q$):
the spectrum of two lower energy HBs as a function of wave vector $(k,k_y)$, calculated at $U=\frac{1}{2}$, $q=100$, $\rho=\frac{1}{a}$ for the first HB a) and $\rho=\frac{4}{a}$ for the second HB b).
}
\label{fig:2}
\end{figure}

\subsection*{Composite Hofstadter bands}

First of all, let's first define what is meant by the term "composite HBs". For this purpose, let us perform numerical calculations of the dispersion of HBs in the semiclassical limit for $\nu=\frac{1}{2}$ filling.
The formation of a $2q-$cell leads to the splitting of HBs into two subbands, and their structure does not depend on $q$ (see Figs.1 a),b)). These subbands intersect at the Dirac point at $k=k_y=\pi$. The dispersion of the spectrum is sharply anisotropic along the direction of the wave vector, so that at $q=20$ it is of order $4\cdot 10^{-2}$ in the $k_y$- direction and $5\cdot 10^{-10}$ in the $k$-direction (see Fig.1c)), at $q=100$ we have the following values $5\cdot 10^{-3}$ and $4 \cdot 10^{-14}$ (see Fig.1d)). In the semiclassical limit, the gap in the excitation spectrum closes and dispersion is transformed into a Dirac point, the spectrum itself is symmetric with respect to the half-filling, which once again confirms the results obtained in \cite{Son}. The calculations of the spectrum of the lowest HB are given in Figs 1, we do not give similar calculations of other bands corresponding to the edge of the spectrum of the model Hamiltonian because their behavior is analogous. 
At $3q$-cell HBs split into three subbands (see Figs 2). In the semiclassical limit the spectrum of quasi-particle excitations is characterized by two Dirac points at $k=\pi$, $k_y=0,\pi$, where each Dirac point is similar to the one at half-filling (see Figs 1 c),d)). In semiclassical limit the lowest and highest subbands in HBs are symmetric about their centrum, at same time the energies of these subbands are shiften on $\pi$ in $k_y$-direction. 

It follows from the numerical results that in the semiclassical limit in a periodic potential with a cell equal to $a=\frac{1}{\nu} q$, the Dirac points split each HB into $\frac{1}{\nu}$-subbands, from which composite HB is formed.

\subsection*{Hall conductance of composite Hofstadter bands}

\begin{figure}[tp]
     \centering{\leavevmode}
     \begin{minipage}[h]{.4915\linewidth}
\center{\includegraphics[width=\linewidth]{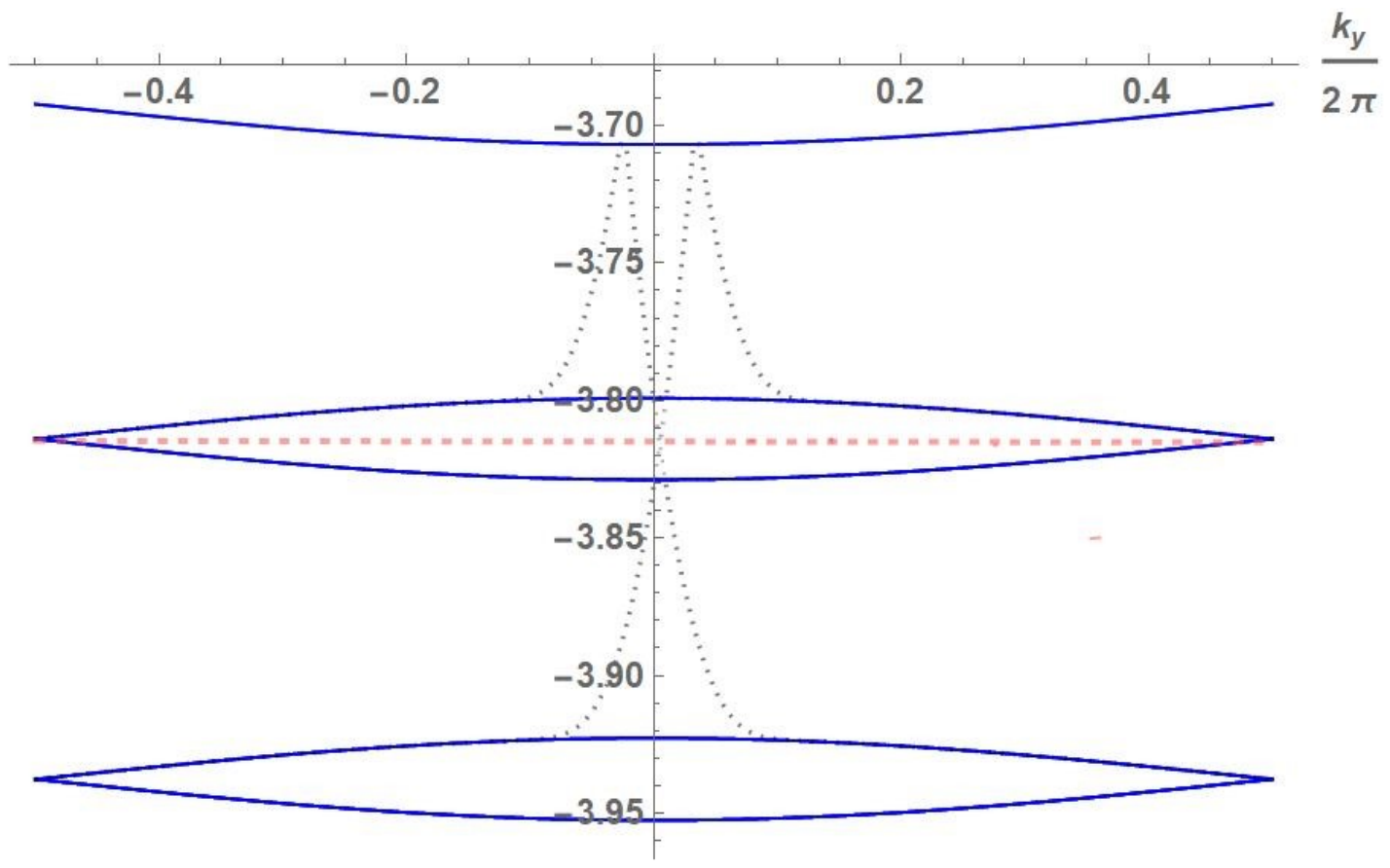} a)\\
               }
    \end{minipage}
     \centering{\leavevmode}
\begin{minipage}[h]{.4915\linewidth}
\center{\includegraphics[width=\linewidth]{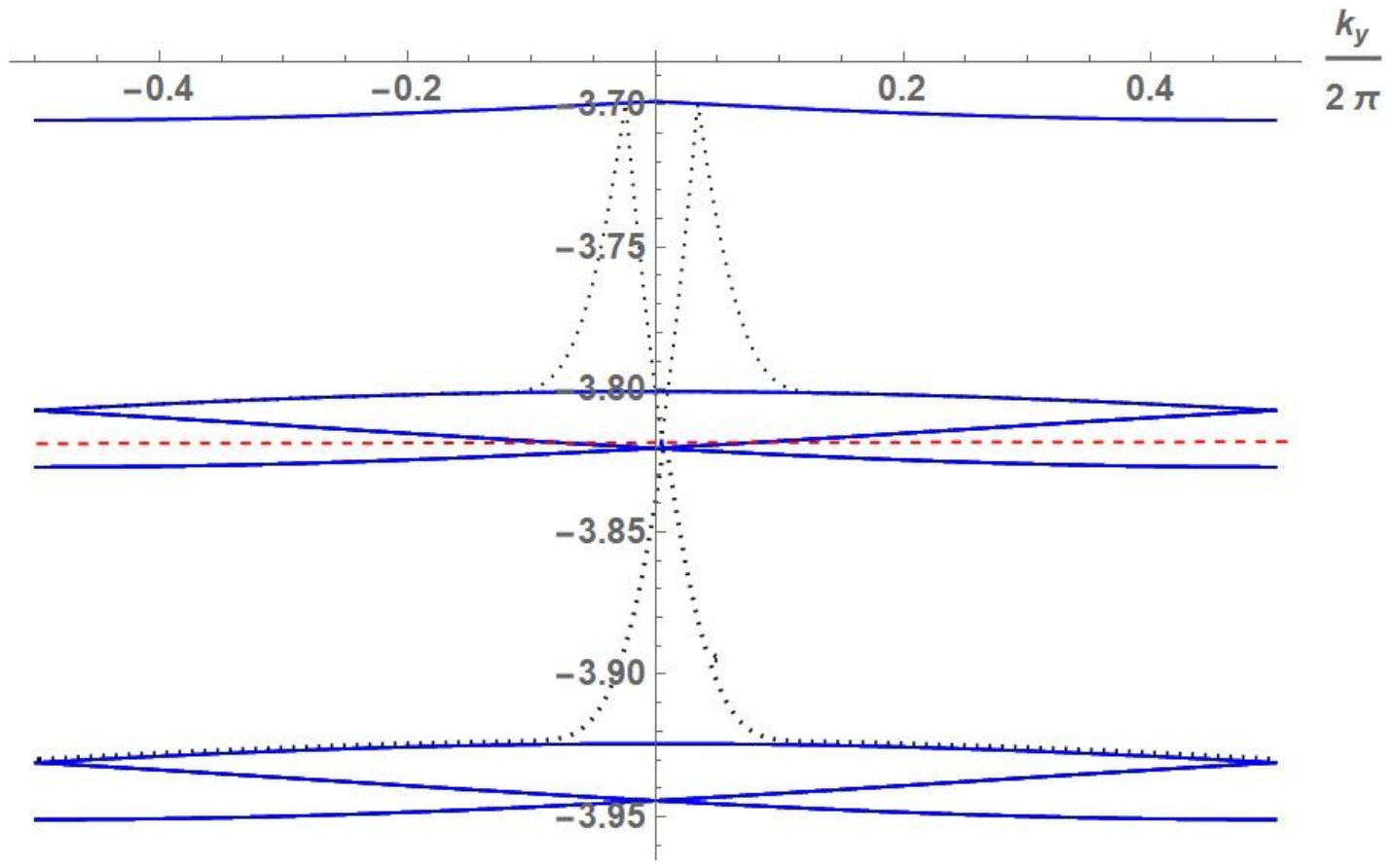} b)\\
                }
    \end{minipage}
\caption{(Color online)\\
$\frac{1}{2}$-filling ($a=2 q$) $\rho=\frac{3}{a}$ a),
$\frac{1}{3}$-filling ($a=3 q$) $\rho=\frac{4}{a}$ b)\\
a fine structure of the spectrum of two lower energy HBs as a function of wave vector $k_y$, calculated at $q=100$, $U=1$ for sample in the form of a hollow cylinder with open boundary conditions along $y$-direction, 
red dashed lines denote the Fermi energies, the dotted lines mark the dispersion of chiral edge modes.
}
\label{fig:3}
\end{figure}

The topological order of $\gamma$-isolated band is denoted by the Chern number $C_{\gamma}$
\begin{eqnarray}
&& C_{\gamma}=\frac{1}{2 \pi}\int_{BZ}{\it B}_{\gamma} (\textbf{k})d^2 k,
\label{eq-4}
\end{eqnarray}
where the Berry curvature ${\it B}_{\gamma} (\bf{k}) =\nabla_{\textbf{k}}\times {\textbf{A}}_{\gamma}(\textbf{k})$ is integrated over the Brillouin zone (BZ). The Berry potential
$$ {\textbf{A}}_{\gamma}(\textbf{k})=-i <u_{\gamma}(\textbf{k})|\nabla_{\textbf{k}}|u_{\gamma}(\textbf{k})> $$
is defined in terms of the Bloch functions $u_{\gamma}(\textbf{k})$. The Chern number for the set of isolated bands is equal to $C_{\Gamma}=\sum_{\gamma \in\Gamma}C_{\gamma}$.

At the point of topological phase transition the Chern number changes, the gap in the spectrum of quasi-particle excitations closes at the Dirac point(s). In the simplest case the Chern number changes by a jump from unity to zero at the phase transition point \cite{IK1}, which corresponds to the Dirac spectrum, the Chern number is equal to half \cite{Hal}. Thus, in the case of a symmetric two-band excitation spectrum in which the subbands are separated by a Dirac point, the Chern number of each subband is equal to $\frac{C}{2}$, where $C$ is the Chern number of an isolated band. Since $\frac{1}{2}$-filling corresponds to a symmetric spectrum of HBs in which two subbands are separated by a Dirac point, there is  $\nu=\frac{1}{2}$ FQHE \cite{Son}.

 As an example, in Fig 3a) we present calculations of the fermion spectrum of the two lowest energy HBs for $\frac{1}{2}$-filling at  $q=100$, $U=\frac{1}{2}$ and $\rho=\frac{3}{a}$ ($a=2 q$). The value of gap between HBs equal to $ 10^{-1}$, the subband width $ 3\cdot 10^{-2}$ and quasi-gap between subbands $ 10^{-14}$ (for comparison).
The periodic potential does not break the time reversal symmetry, so its addition to the Hofstadter model does not change the Chern numbers, which characterize the topological state of the fermion liquid. As expected, the number of chiral modes localized at the sample boundaries does not depend on the periodic potential, the Chern numbers of each HB are equal to unity (see Fig 3 a)). The spectrum of each HB is symmetric about the band centrum, $\frac{1}{2}$-filling each bands corresponds to Hall conductance with the Chern number $\frac{1}{2}$. Taking into account $\gamma$ isolated full-filled HBs,  we obtain   $C_{\Gamma}=\sum_{\gamma \in\Gamma}C_{\gamma}+\frac{1}{2}$. Thus the numerical calculations obtained within the Hamiltonian (1),(3) agree with \cite{Son}. 

Let us calculate the spectrum of two lower energy HBs for $\frac{1}{3}$-filling of the second band, the calculations are illustrated in Fig 3b). As we show below, the behavior of fermion liquid for an arbitrary fractional filling is no different from $\frac{1}{2}$ discussed above. Since in the semiclassical limit  the spectra of HBs are symmetric about their center (in this case, the spectra of the outermost subbands  are shifted by $\pi$ relative to each other). The Chern number of each subband is equal to $\frac{1}{3}$, as in the case of $\frac{1}{2}$-filling, since the Chern number for a separate HB is equal to one. 

Thus at $\nu$-filling of $\gamma+1$ HB, the Chern number, which determines the Hall conductance is equal to
 $C_{\Gamma}=\sum_{\gamma \in\Gamma}C_{\gamma}+\nu C_{\gamma+1}$.

\section*{Conclusions}

In the semiclassical limit (which satisfies experimental conditions) within the Hofstadter model
the Chern number was calculated for fractional filling of HB. Numerical calculations have shown that the periodic potential can lead to a steady state with lattice formation with the same period \cite{IK0}. In this case a fine structure of the spectrum of HBs  is formed with the formation of subbands, the subbands are separated by the Dirac points (it is very important). 
In other words, composite HBs are formed. The Chern number of a fully filled $\gamma$-composite HB does not differ from that of HB $C_{\gamma}$.  Due to the symmetry of the spectrum of quasi-particle excitations of composite HBs relative to their center, at $\nu$-filling of $\gamma$-composite HB its Chern number is equal to $\nu C_{\gamma}$. 
The results are in agreement with calculations for $\frac{1}{2}$-filling \cite{Son}, in which the composite fermion paradigm is taken as a basis. To the question of Dam Son ”Is a composite fermion
a Dirac particle?” \cite{Son} can be answered in the affirmative in the sense that HB, which defines the spectrum
of quasiparticle excitations, is a composite one. 
Thus, within the framework of a unified approach, the nature of both integer and fractional quantum Hall effects, which are directly realized on the experiment, is studied. This allows us to consistently describe the experiment rather than to consider separate sequences of the Hall plateaus \cite{Son2}.

 \section*{Author contributions statement} I.K. is an author of the manuscript

\section*{Additional information} The author declares no competing financial interests. \\

\section*{Availability of Data and Materials}
Data generated or analyzed during this study are included in this published article.\\
All data generated during the study are available from corresponding author on reasonable request.\\
Correspondence and requests for materials should be addressed to I.K.
\end{document}